\documentclass[aps,prl,onecolumn,groupedaddress,showpacs]{revtex4}
\usepackage{amsmath}
\usepackage{graphicx}

\begin{document}

\title{Sound modes in composite incommensurate crystals.}

\author{R. Currat}
\affiliation {Laue-Langevin Institute, F-38042, Grenoble, France }

\author{E. Kats}
\affiliation {Laue-Langevin Institute, F-38042, Grenoble, France }
\affiliation {L. D. Landau Institute for Theoretical Physics, RAS,
117940 GSP-1, Moscow, Russia}

\author{I. Luk'yanchuk}
\affiliation {Laue-Langevin Institute, F-38042, Grenoble, France }
\affiliation {L. D. Landau Institute for Theoretical Physics, RAS, 117940 GSP-1, Moscow, Russia}

\date{\today}

\begin{abstract}
We propose a simple phenomenological model describing composite crystals,
constructed from two parallel sets of periodic inter-penetrating chains. 
In the harmonic approximation and
neglecting thermal fluctuations we find the eigenmodes of the system. It is
shown that at high frequencies there are two longitudinal sound modes with
standard attenuation, while in the low frequency region there is one
propagating sound mode and an over-damped phase mode. The crossover between
these two regions is analyzed numerically and the dynamical structure factor
is calculated. It is shown that the qualitative features of the experimentally observed
spectra can be consistently described by our model.
\end{abstract}


\maketitle

\section{I. Introduction}

Among the various types of incommensurate systems one of the simplest kind (the so-called 
uniaxial composite crystal) is constructed from 
two sets of regular inter-penetrating incommensurate chains. The
recurrent interest for these systems is related mainly with the experimental activity on
several classes of intergrowth compounds which can be viewed as good physical realizations of the 
uniaxial 
composite crystal. There is a considerable literature (mainly theoretical but not only) discussing
eigenmodes and related properties of composite systems (see e.g. the review article \cite{CJ88} and
references therein, and the more recent papers quoted in \cite{RJ97}, \cite{RJ99}).
 
Although a number of sophisticated calculations have been published
over the last 20 years
\cite{TR78,EA78,FR82,FR83,AB82,BP83}, there is still a clear need for a simple (but yet
non-trivial) theoretical
model with predictions which can be directly tested experimentally. In previous models 
all branches of excitations were simultaneously examined \cite{BP83}, and
many subtle details of the composite systems were discussed in connection with higher order
commensurability effects \cite{TR78},
such as the hierarchical nature of gaps arising at commensurate -
incommensurate phase transitions \cite{RJ97}.
The results of these investigations, even in the long wavelength limit, are expressed in terms of many
unknown parameters, and lead to many different modes, 
while experimentally it is not clear whether it is possible at all 
to observe
these modes and to determine these parameters. 
Furthermore, although the basic theory of excitations in
incommensurate and composite systems was born long ago \cite{FK38}, \cite{OV71}, \cite{AX80}, many
questions of
principle remain unsettled, and there is an impressive quantity of unexplained, partially explained or
contradictory results. In part this frustrating situation is just due to the lack of a simple and
tractable analytical model.
 
Our motivation for adding one more paper to the topic is precisely to propose such a model. We do not
intend to discuss in the present paper any of the dynamical details connected with the description of
discontinuity edges in the vibrational density of states nor do we deal with 
the many structural properties
(see e.g. \cite{RJ97}) related to commensurate - incommensurate phase transitions. Our aim is much more
modest and based on the fact that many robust and experimentally testable features of the excitation
spectra are not sensitive to the more delicate aspects
of commensurability effects.     

Note that one can consider composite structures as a new state of
matter like liquid or solid states.
Composite structures differ from
modulated crystals insofar as one can not define an average periodic
structure for the full system. Unlike crystals with incommensurate density
waves that are characterized by weak deviations of atoms relative to
their regular positions and unlike the adsorption structures with the rigid
host and flexible guest sublattices, both subsystems in composites are more
or less equivalent and therefore one can not treat them perturbatively as an
external potential acting on its elastic counterpart, nor can one 
use the classical Frenkel -
Kontorova model \cite{FK38} to describe phase diagrams and excitations.
Note that composite incommensurate crystals are also different from 
quasicrystals.
According to the traditional classification \cite{CL95} 
quasicrystals refer to systems where the rotational symmetry 
corresponds
to a forbidden crystal symmetry, which determines the unique ratio of
incommensurate length scales that defines the structure, whereas for incommensurate
crystals the rotational symmetry corresponds to an allowed crystal symmetry
and many different sets of incommensurate length scales are possible. 

Due to the existence
of (at least) two length scales, there are twice as many hydrodynamic variables as
in conventional crystals. However in quasicrystals these additional
variables cannot be interpreted as real displacements of any of the sublattices,
and this distinction leads to 
specific physical consequences.

\begin{figure}[!h]
\centerline{\includegraphics[width=6cm]{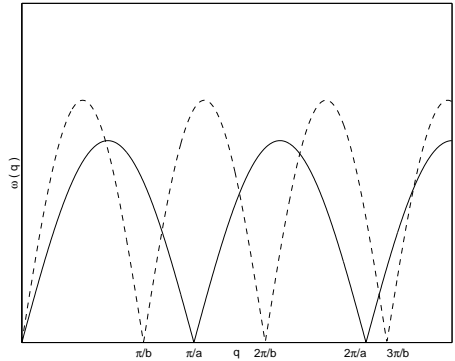}}
\caption{Phonon spectra of two uncoupled incommensurate chains $a$ and $b$.}
\label{twobranch}
\end{figure}

The most interesting aspect of the physics of composites is the dynamics of
their sublattices. In order to formulate the problem, 
let us consider first two inter-penetrating
atomic chains with periods $a$ and $b$ and assume no coupling between them.
Then, two independent sets of incommensurate (in $q$-space) phonon branches
with dispersions $\omega _{a}(q)=c_{a}a^{-1}\left| \sin qa\right| $ and $%
\omega _{b}(q)=c_{b}b^{-1}\left| \sin qb\right| $ should be observed (Fig.%
\ref{twobranch}). What happens with them, when the inter-chain interaction is
properly taken into account? Different types of coupling are responsible
for different effects.

\textit{The elastic coupling} results in the renormalization of the  sound
velocities \cite{AB82}. It leads to repulsion of the crossing branches with
formation of intermediate gaps. The umklapp intersection of branches occurs
however for \textit{any} $q$ since, because of the period
incommensurability, the condition $\omega _{a}(Nq)\approx \omega _{b}(Mq)$
can always be satisfied when the integers $N,M$ are chosen properly. Hence a
rich hierarchical set of gaps in phonon spectra is expected (see e.g. \cite{RJ97}).
Some general features related to breaking of analyticity due to
static elastic couplings in the 1d double-chain model were reported recently
in \cite{RJ99}.

\textit{The electrostatic coupling} takes place when the chains $a$ and $ b $ 
are charged. It leads to nonlocal renormalization of the
phonon self-energy, nontrivial retardation effects,
and to the plasmon gap formation \cite{FR82}, \cite{FR83}  \cite{AB82}.

\textit{The dynamical dissipative coupling } originates from the internal
friction between the chains. To our knowledge the dissipative dynamics of
composites was studied only by the authors in Refs. 
\cite{FR82,FR83} and \cite{BP83}.
In the papers \cite{FR82,FR83} the authors have taken into account
dynamic resonance processes arising due to the existence of degeneracy
points of the subsystem modes which allow the interchange of energy
and momentum. They described these processes by
including into the response matrix a finite (in the limit $q \to 0$,
and $\omega \to 0$) off-diagonal contribution. For a strictly
harmonic lattice the contribution vanishes identically.
We propose below
a microscopic realization of the dissipative dynamical coupling which is
always present in real systems even in harmonic approximation.
As it concerns to the paper \cite{BP83} the authors took into 
consideration all possible types of couplings and all degrees of freedom
of composite systems, and therefore it is difficult to
survey their results (and all the more to compare the results
with experimental data) because of the large number of hydrodynamic
variables involved.
Meanwhile the broadening and renormalization of phonon spectra due to the
inter-chain friction can easily hide the subtle hierarchical gap structure as
well as other minute effects, like the fluctuation - driven destruction of the 1d
composite long - range order \cite{EA78} or 
the static and dynamic lock-in effects \cite{RJ97}.

It is the purpose of the present publication to provide a simple and
tractable model that incorporates the dissipative dynamic coupling in composite
systems to calculate the experimentally testable features of the excitation
spectra.
We put forward a model of a composite system, 
which can be used to find the eigenmode spectrum 
(under relatively weak assumptions), 
and to compute the dynamical structure factor. 

The outline of this article is as follows. In Section II we formulate a
model description of a composite incommensurate crystal which includes all the basic dynamical
equations. In Sections III and IV 
the calculations of the eigenmodes and dynamical structure factors are presented.
Section V is devoted to a discussion and summary of our main results. The Appendix  
deals with a very schematic description
of a particular microscopic model of a uniaxial composite system.                                                    

\section{II. Model} \label{sec:Model} 

The simplest composite crystal consists of alternating chains with
incommensurate periods $a$ and $b$ that are located in the $x - y$ plane and
are directed along axes parallel to $x$ at $y=md$ and at $y=(m+1/2)d$. 
In this case even the basic structure is not commensurate. Because of the
local differences in environment the $a$ chain is modulated with period
$b$, and the $b$ chain - with period $a$. Therefore the 
atomic positions can be given by 
\begin{eqnarray}
\mathbf{R}^{a}(n , m ; t)=md\mathbf{e}_{y}+na\mathbf{e}_{x}+\mathbf{u}^{a}(n , m ; t) + 
\notag \\
{\bf e}_x g_1(na - \Delta )  \,, (n,m=0,\pm 1,...)  
\, ;
\notag \\
\mathbf{R}^{b}(n , m ; t)=(m+1/2)d\mathbf{e}_{y}+nb\mathbf{e}_{x}+\mathbf{u}^{b}(n , m ; t) 
+ \Delta +
\notag \\
{\bf e}_x g_2(mb + \Delta ) \, ,
\label{ckl1}
\end{eqnarray}
where $\Delta $ is a certain natural offset of one set of chains with
respect to its counterpart, ${\bf u}^{a,b}(n , m ; t)$ are the
displacements of the atoms about
their average equilibrium positions, 
$g_1(x + b) \equiv g_1(x)$ and $g_2(x+a) \equiv g_2(x)$ are modulation
functions.

In fact just the definitions (\ref{ckl1}) allow us to introduce
all the degrees of freedom 
relevant for our model 
of a composite system
(including therefore - the main branches of excitations). If we neglect
for a moment the interchain interaction (purely hypothetically since
in the absence of subsystem $b$ subsystem $a$ would be generally unstable),
we get evidently two zero - energy modes for the two chains, each
one implying a rigid displacement of one chain with respect to its
fixed counterpart.

Interchain interactions introduce new ingredients in this scheme.
In the incommensurate state the position of the centre of mass of
the atoms relative to the modulation potential can be described by a phase
variable. In a rigid chain approximation the acoustic phonon mode
corresponds to a global rigid displacement of the two sets of chains,
while the phase mode corresponds
to relative rigid displacements of the
two chains in opposite directions such that the centre of mass of the
composite crystal remains fixed.
In a commensurate state this phase variable is pinned, thus there is
one acoustical mode and one optical mode with a gap whose magnitude
is determined by the pinning potential and strongly decays with
the order of commensurability. 
In the latter case our consideration will be still valid for 
characteristic frequencies
higher than the gap.

Considering the relative displacements to be rigid means that
we neglect the above mentioned intermodulation. The interchain interaction
moves the atoms on both chains away from their regularly distributed positions.
However if the modulation functions $g_1$ and $g_2$ are differentiable,
then the atomic displacements corresponding to the phase mode are given
by the following ''dressed'' displacements $\tilde {{\bf u}}$
\begin{eqnarray}
\label{phason}
\tilde {{\bf u}}^a(n , m ; t) \to {\bf u}^a(n , m ; t) + {\bf e}_x b g_1^\prime \, , \,
\tilde {{\bf u}}^b(n , m ; t) \to {\bf u}^b(n , m ; t) + {\bf e}_x a g_2^\prime \, ,
\end{eqnarray}
where $g_i^\prime $ denote a derivative over the argument.

In order to discuss the long-wavelength, low-frequency excitations
one should transform the discrete
variables 
introduced above into
continuous ones.
For sound modes which consist of equal shifts of
$\mathbf{R}^{a}(n , m ; t)$ and $\mathbf{R}^{b}(n , m ; t)$ this transformation
is evident. Namely in a continuum approximation the discrete variables $\mathbf{u}^{a,b}(n , m ; t) $ 
are substituted by continuous functions 
$\mathbf{u}^{a,b}(\mathbf{r},t)$.
However it is not the case for the phase mode which consists 
of inhomogeneous (on a microscopic scale) atomic displacements 
(corresponding to sliding
of one chain system with respect to the other). However when the modulation functions
$g_i$ are analytic, one can still use the long-wavelength description but
for dressed displacements ${\tilde {\bf u}}$.

For the benefit of the more skeptical reader the question concerning ''dressed''
variables should be clarified in more detail. In fact the dressed displacements
${\tilde {\bf u}}$ depend on the displacements ${\bf u}^a$ and ${\bf u}^b$ 
as  
initially defined according to
(\ref{ckl1}) 
(since the
modulation functions depend on the difference $u_x^a - u_x^b$).
For weak enough interactions and far enough away from commensurability,
the modulation functions $g_{1,2}$ are continuous analytic functions and the energy 
of the system remains unchanged when the phase $u_x^a - u_x^b$ is varied.
However a uniform shift $u_x^a - u_x^b$ does not indicate a uniform displacement
of the atoms in both chains, since the functions $g_1$ 
and $g_2$ are not constants.
Thus strictly speaking we should write down equations of motion
separately for dressed and undressed variables. In our phenomenological
approach it means a certain renormalization of all phenomenological
constants entering the equations.
In what follows we neglect the renormalization and 
will use 
$\mathbf{u}^{a,b}(\mathbf{r},t)$ for both (acoustical and phase)
modes.
In spite of this erroneous assumption (which is equivalent
to $g_{1 , 2} \simeq const$) the approximation correctly identifies
the important modes and characteristic scales in the problem. 
This sin of omission can be easily corrected when detailed information
concerning the values of the parameters becomes available.

The elastic energy can be written 
as a quadratic form of the strain tensor $\epsilon _{ij}^{\alpha }=
\frac{1}{2}(\partial _{i}u_{j}^{\alpha }+\partial _{j}u_{i}^{\alpha })\,$: 
\begin{eqnarray}
\mathcal{F} &=&\frac{1}{2}\lambda _{ijkl}^{\alpha \beta }\epsilon _{ij}^{\alpha
}\epsilon _{kl}^{\beta }+\frac{1}{2}\sigma (u_{y}^{a}-u_{y}^{b})^{2};
\label{ckl2} \\
\alpha ,\beta &=&a,b;\quad i,j,k,l=x,y\,.  \notag
\end{eqnarray}
The inter-chain interaction is given by the relative displacements
along $y$ (the 2nd term in (\ref{ckl2})) and also by the cross terms in the
elasticity tensor $\lambda _{ijkl}^{ab}$. At the same time the free energy
expansion (\ref{ckl2}) is
invariant with respect to the relative shift of the chains along $x$ because
of the incommensurability of the two periods, therefore there is no
terms depending explicitely on $u_x^{a , b}$ (only on gradients of
the displacements, i.e. on $\epsilon _{i j}^{a , b}$).

As a note of caution we should remark here that there is
a fundamental difference between the both kind of soft degrees
of freedom in (\ref{ckl2}), i.e. between acoustic and phase modes.
The first one is a true Goldstone mode, because it breaks the continuous
translational symmetry (and therefore remains hydrodynamic
and gapless even if the modulation functions are not continuous).
But the latter is a pseudo-Goldstone mode and it loses its hydrodynamic
character in the case of discontinuous modulation functions (since phase
variable breaks
only countable translational symmetry).

To complete the derivation of the dynamical equations we have to know not
only the elastic energy (\ref{ckl2}) but also the dissipative function $R$.
The internal friction produced by the sliding displacement of the chains
results in a dynamical inter-chain coupling proportional to the chain
velocity difference $\overset{.}{u}_{x}^{a}-\overset{.}{u}_{x}^{b} $ 
that contributes to the dissipation function: 
\begin{equation}
R=\frac{1}{2}\Gamma (\overset{.}{u}_{x}^{a}-\overset{.}{u}_{x}^{b})^{2}
+\frac{1}{2}\eta _{ijkl}^{\alpha \beta }\overset{.}{\epsilon }_{ij}^{\alpha }
\overset{.}{\epsilon }_{kl}^{\beta }\,.
\label{ckl3}
\end{equation}
The second term in (\ref{ckl3}) corresponds to the viscous dissipation, and
the coefficients $\eta _{ijkl}^{\alpha \beta }$ entering (\ref{ckl3}) have
the standard meaning of viscosity tensor elements \cite{LL}. One new
phenomenological coefficient $\Gamma $ describes a friction proportional to
the velocity difference between the two chains.
A few comments to assign physical meanings to the terms in
equation (\ref{ckl3}) may be helpful to the reader.
The first term in (\ref{ckl3}) is the energy dissipation caused by relative 
motion between the two set of chains.
In the framework of our purely phenomenological approach we are not in a
position to determine a dominant microscopic mechanism for the dissipation.
A natural estimate for the order of magnitude of the friction coefficient
$\Gamma $ (based on the assumption that the work produced by
the drag force density is completely dissipated in a viscous media) is
$\Gamma \propto \eta /d^2$, where $\eta $ is the characteristic viscosity.
It is worth noting that dynamic dissipative, static
elastic and lock-in interchain interactions can 
differ both quantitatively, since there
is no simple relation between the coupling coefficients $\Gamma $, $\sigma $,
and the elastic moduli, and qualitatively, because unlike static interactions,
the dynamic dissipative coupling does
not necessarily produces intermodulations
of the two subsystems.

Now we are in a position to write down the equations of motion 
\begin{gather}
\rho _{\alpha \beta }\overset{..}{u}_{i}^{\alpha }=-\frac{\partial 
\mathcal{F}}{\partial u_{i}^{\alpha }}-\frac{\partial R}{\partial 
\overset{.}{u}_{i}^{\alpha }} \label{Newton}\\
=\lambda _{ijkl}^{\alpha \beta }\partial _{j}\epsilon _{kl}^{\beta }+\eta
_{ijkl}^{\alpha \beta }\partial _{j}\overset{.}{\epsilon }_{kl}^{\beta } + \Gamma
\delta _{ix}(\delta ^{\alpha a} - \delta ^{\alpha b})(\overset{.}{u}_{i}^{a}-\overset{.}{u}_{i}^{b})
+ \sigma \delta _{iy}(\delta ^{\alpha a} - \delta ^{\alpha b})(u_{i}^{a}-u_{i}^{b})\,,  \notag
\end{gather}
where 
\begin{equation}
\rho _{\alpha \beta }=\left( 
\begin{array}{cc}
\rho _{a} & 0 \\ 
0 & \rho _{b}
\end{array}
\right)
\end{equation}
is the matrix of densities. Eq. (\ref{Newton}) is the second Newton law where
the forces acting on the component $a$ are: (i) the friction between the
components due to their relative motion, and (ii) the elastic forces.

As it is well known \cite{LL} the propagation of elastic waves in 
anisotropic media is a rather complicated phenomenon. The directions
of polarization and the eigen-frequencies are determined by the
dispersion equation which is derived for the monochromatic
solutions of (\ref{Newton}). In a general anisotropic case
however none of the polarization directions correspond either
purely longitudinal or purely transverse directions with respect to
the direction of the wave vector. 
We can look for solutions of the
solution of the equations of motion (\ref{Newton}) in the form
\begin{equation}
u_{i}^{\alpha } = u_{i0}^{\alpha }e^{i(\mathbf{qr}-\omega t)}.
\label{wave}
\end{equation}
The dispersion $\omega (\mathbf{q)}$ can be found by substitution of (\ref
{wave}) into (\ref{Newton}) and, then, by diagonalization of the corresponding
dynamical equations: 
\begin{equation}
D_{\alpha \beta }^{ij}(q,\omega )u_{i0}^{\alpha }=0.  \label{de}
\end{equation}
The four degrees of freedom of $u_{i}^{\alpha }$ ($\alpha =a,b$, $i=x,y$)
correspond to the four propagating eigenmodes.

We are interesting in the case when the excited wave is propagating along 
$\mathbf{x}$ (other geometries will be shortly discussed in the Conclusion). Then,
the equation of motion (\ref{Newton}) can be split into a transverse part for 
$u_{y}^{\alpha }$ and a longitudinal part for $u_{x}^{\alpha }$. In the
transverse case, the last term in Eq. (\ref{Newton}) results in the
decoupling of the excitations into sound and optical modes. 
Thus for the polarization perpendicular to the chains we will find
the usual optical and acoustical branches.
Less trivial is
the longitudinal case when the last term in Eq. (\ref{Newton}) vanishes and
the dynamical mode coupling $\Gamma \delta _{ix}(\overset{.}{u}_{i}^{\alpha
}-\overset{.}{u}_{i}^{\beta })$ provides the most important contribution to
the properties of the propagating excitations. In the next Section we
consider this case in detail.

\section{III. Longitudinal excitation spectrum}\label{sec:Spectrum}

The dynamical equation for the longitudinal eigenmode with 
$\mathbf{q\parallel x}$ looks like 
\begin{equation}
D_{\alpha \beta }^{xx}(q,\omega )u_{x0}^{\alpha }=0\,  \label{disp}
\end{equation}
with 
\begin{widetext}
\begin{equation}
D_{\alpha \beta }^{xx}(q,\omega )=\left( 
\begin{array}{cc}
-\rho _{\alpha }\omega ^{2}+\lambda _{xxxx}^{aa}q^{2}-i\Gamma \omega -i\eta
_{xxxx}^{aa}q^{2}\omega & \lambda _{xxxx} ^{ab}q^{2}+i\Gamma \omega -i\eta
_{xxxx}^{ab}q^{2}\omega \\ 
\lambda _{xxxx}^{ab}q^{2}+i\Gamma \omega -i\eta _{xxxx}^{ab}q^{2}\omega & 
-\rho _{b}\omega ^{2}+\lambda _{xxxx}^{bb}q^{2}-i\Gamma \omega -i\eta
_{xxxx}^{bb}q^{2}\omega
\end{array}
\right) .  \label{matr}
\end{equation}
\end{widetext}

The terms with $\Gamma $ and $\lambda ^{ab}$ provide the dynamical
dissipative and elastic coupling between two sound waves propagating in
chains $a$ and $b$. To pick out the effect of this coupling on the phonon
dispersions we neglect the intrinsic phonon attenuation that is given by
terms with $\eta $.

To find the eigenmodes one should solve the characteristic equation: 
\begin{eqnarray}
\det D_{\alpha \beta }^{xx}(q,\omega ) &=&\rho _{\alpha }\rho _{b}[(\omega
^{2}-c_{a}^{2}q^{2}+i\gamma _{a}\omega )(\omega ^{2}-c_{b}^{2}q^{2}
\notag \\
+i\gamma _{b}\omega ) &-&(c_{i}^{2}q^{2}+i(\gamma _{a}\gamma
_{b})^{1/2}\omega )^{2}]=0
\label{char} 
\end{eqnarray}
where:

\begin{gather}
c_{a}^{2}=\lambda _{xxxx}^{aa}/\rho _{\alpha },\quad c_{b}^{2}=\lambda
_{xxxx}^{bb}/\rho _{b},\quad c_{i}^{2}=\lambda _{xxxx}^{ab}/(\rho _{\alpha
}\rho _{b})^{1/2},  \notag \\
\gamma _{a}=\Gamma /\rho _{\alpha },\quad \gamma _{b}=\Gamma /\rho _{b},
\end{gather}

\textbf{At small} $q$ ($q<\gamma _{a,b}/c_{a,b}$) the dynamical
dissipative coupling substantially renormalizes the dispersions of
the two sound modes $a$ and $b$.

The first two branches correspond to the time-conjugated sound-like modes 
\begin{equation}
\label{x}
\omega (q)\approx \pm c_{s}q-i\eta _{s}q^{2}\,,
\end{equation}
with effective velocity and damping: 
\begin{gather}
c_{s}^{2}=\frac{\gamma _{a}c_{b}^{2}+\gamma _{b}c_{a}^{2}-2\left( \gamma
_{a}\gamma _{b}\right) ^{1/2}c_{i}^{2}}{\gamma _{a}+\gamma _{b}}, \\
\eta _{s}=\frac{1}{2\left( \gamma _{a}+\gamma _{b}\right) ^{2}}\frac{\left[
\left( \gamma _{a}-\gamma _{b}\right) c_{i}^{2}-\left( \gamma _{a}\gamma
_{b}\right) ^{1/2}\left( c_{a}^{2}-c_{b}^{2}\right) \right] ^{2}}{\gamma
_{a}c_{b}^{2}+\gamma _{b}c_{a}^{2}-2\left( \gamma _{a}\gamma _{b}\right)
^{1/2}c_{i}^{2}}.  \notag
\end{gather}

The third mode corresponds to the over-damped phason-like mode with pure
imaginary dispersion (diffusive mode): 
\begin{equation}
\label{x1}
\omega \approx -i\eta _{ph}q^{2},
\end{equation}
where the damping factor is given by:

\begin{equation}
\eta _{ph}=-\frac{c_{a}^{2}c_{b}^{2}-c_{i}^{2}}{\gamma _{a}c_{b}^{2}+\gamma
_{b}c_{a}^{2}-2\left( \gamma _{a}\gamma _{b}\right) ^{1/2}c_{i}^{2}}.
\end{equation}

In addition there is a purely relaxational fourth mode: 
\begin{equation}
\label{x2}
\omega =-i(\gamma _{a}+\gamma _{b})
\end{equation}
which, however, is hidden by the third mode that relaxes much more
slowly.

\textbf{At high} $q$: ($q>\gamma _{a,b}/c_{a,b}$) the elastic coupling leads
to the effective renormalization of the velocities of the sound modes $a$ and $b$.
The effect of the dynamical (dissipative) coupling is manifested by a specific $q$
-independent sound attenuation. The dispersion of these two
time-conjugated sound modes reads: 
\begin{equation}
\label{x3}
\omega =\pm c_{12}q-i\gamma _{12}.
\end{equation}
The renormalized velocities and attenuation factor are given by:

\begin{equation}
2c_{12}^{2}=c_{a}^{2}(1\pm \kappa )+c_{b}^{2}(1\mp \kappa ),
\end{equation}
\begin{equation}
4\gamma _{12}=\left( \sqrt{\gamma _{a}(1\pm \kappa ^{-1})}-\sqrt{\gamma
_{b}(1\mp \kappa ^{-1})}\right) ^{2}.
\end{equation}
where the elastic coupling parameter $\kappa $: 
\begin{equation}
\kappa =\sqrt{1+\frac{4c_{i}^{4}}{(c_{a}^{2}-c_{b}^{2})^{2}}}
\end{equation}
is larger than one (weak coupling) and smaller than $(c_{a}^{2}+c_{b}^{2})/%
\left| c_{a}^{2}-c_{b}^{2}\right| $ (stability condition).

At higher $q$ the intrinsic phonon attenuation $\sim -i(\eta
_{xxxx}^{aa,bb}/\rho _{(a , b)})q^{2}$ is superimposed onto the dynamical
attenuation $-i\gamma _{12}$, a situation which
can lead to a nontrivial $q$-dependence
of the phonon line-width.

Of course it is possible to find the eigenmodes (at least no problem at all to find the dispersion laws
numerically) in the general case of arbitrary $q$ 
taking into account all phenomenological parameters
entering
(\ref{matr}). However aiming for a simple model we have chosen 
the simplest coupling form for the
displacement components on both chains, following the principle 
of minimal requirements. To keep our model
tractable (but yet non-trivial) and to avoid tedious 
calculations we drop the intra-mode coupling 
$\lambda _{xxxx}^{ab}q^{2}$ that does not change the qualitative picture. 
As will be discussed further,
the intrinsic
phonon attenuation expressed by $i\eta _{xxxx}^{\alpha \beta }q^{2}\omega $ can be 
also neglected in the low energy region which is relevant to 
this discussion. Although actually this picture is not entirely correct
it correctly identifies the important modes and characteristic scales in the problem.                

As a note of caution we should also keep in mind that there are four types of 
interactions involved, namely:
\begin{itemize}
\item
the elastic interaction proportional to gradients of the displacements, and related
to off-diagonal terms of the elasticity tensor;
\item
locking interaction, proportional to sublattice displacement differences
(the last term in (\ref{Newton}) with the coefficient $\sigma $);
\item
two dissipative couplings described by the off - diagonal terms
of the viscosity tensor and by the mutual friction coefficient $\Gamma $ in (\ref{Newton}).
\end{itemize}

The most robust phenomena are related to the dissipative couplings which
we have taken
into account in the derivation of the dispersion laws (\ref{x} - \ref{x3}). 
However the locking 
and elastic couplings could lead to appreciable effects, like e.g. a gap at $q=0$ for the
sliding mode.
 
\section{IV. Dynamical structure factor: scattering experiments}\label{sec:Factor}

The above discussed distinctive features of the excitation spectrum should
be observable in different kinds of scattering experiments: ultrasonic
measurements, Brillouin light scattering and inelastic coherent scattering
of neutrons or X-rays. 
The double - differential cross-section gives the probability that say a neutron
with initial energy $E$ is scattered (coherently or incoherently) into
a detector subtending a solid angle $d \Omega $ around a certain direction
with spherical coordinates $2\theta $ and $\phi $, with a final energy
$E^\prime $. We note $\hbar \omega = E - E^\prime $, 
($\hbar \omega $ is the energy transfer), 
and 
$\hbar \mathbf{Q}=\hbar \mathbf{k}-\hbar \mathbf{k}^{\prime }$ is the 
momentum transfer, where $\hbar {\bf k}$ and $\hbar {\bf k}^\prime $,
are the incoming and outgoing neutron momenta. 
The scattered intensity is proportional to the differential cross
section for inelastic scattering: 
\begin{widetext}
\begin{eqnarray}
\frac{d^{2}\sigma }{d\Omega d\omega }=\frac{k^{\prime }}{k}\frac{1}{2\pi
\hbar }\int dte^{i\omega (t-t^{\prime })}
\left \langle \sum_{n,n^{\prime }, m, m^\prime , \alpha ,\beta
}s_{\alpha }s_{\beta } e^{-i\mathbf{QR}^{\alpha
}(n , m ; t)}e^{i\mathbf{QR}^{\beta }(n^\prime , m^\prime ; t^{\prime })} \right \rangle
\,,
\label{f}
\end{eqnarray}
\end{widetext}
where angular brackets denote a thermal average,
$s_{\alpha }=s_{a},s_{b}$ are the coherent atomic
scattering factors
i.e. differential cross sections of light, sound, neutrons or X-rays 
for single atoms of chains $a$ and $b$, 
and the integrand in (\ref{f}) is called the dynamical
structure factor $S({\bf Q} ,\omega )$ which is accessible by scattering techniques,
and obeys the so-called detailed
balance condition
$$
S({\bf Q} , \omega ) = \exp\left (\frac{\hbar \omega }{T}\right ) S(-{\bf Q} ,- \omega ) \, ,
$$
if the sample is in thermal equilibrium.

According to the fluctuation dissipation theorem \cite{LLS} 
$$
S({\bf Q} , \omega ) = \frac{1}{1- \exp(-\hbar \omega /T)}\frac{1}{\pi } \chi ^{\prime 
\prime }({\bf Q} , \omega ) \, ,
$$
where $\chi ^{\prime \prime }$
is  the imaginary part of the generalized susceptibility. 

Calculations of $\chi ^{\prime \prime }$ is a formidable task that can be done
only on the basis of an appropriate microscopic model (see e.g. the Appendix
to our paper). In a general case the dynamic susceptibility $\chi $
is a matrix with respect to the displacement fields of the two chains
and to the respective Cartesian components. Owing to the lack of discrete
translational invariance in the $x$ - direction
it is non-diagonal also in the wave vectors.
Since the full investigation of 2d (all the more 3d) problem
requires an extraordinary amounts of computational work we
must defer the investigation of the full problem to the time
when experimental results 
suitable for a quantitative comparison 
become available (see also the Conclusion section of the
paper), and present here only the results for the wave vector
parallel to the incommensurate direction ($x$ - axis).

In the case of longitudinal phonons it can be written in the following form
(we dropped here unessential factors,
since we are not interested in the absolute value of $S({\bf Q} ,
\omega )$
or $\chi ^{\prime \prime }({\bf Q} , \omega )$): 
\begin{equation}
S(Q,\omega )\propto Re \sum_{n,n^{\prime }, \alpha ,\beta }s_{\alpha
}s_{\beta }e^{iQ(\alpha n-\beta n^{\prime })}Q^{2}\frac{T}{\omega }
Im \chi
_{\alpha \beta }^{xx}(n,n^{\prime },\omega ).  \label{SQ}
\end{equation}
To study the structure factor we need to know
the equilibrium positions of the atoms, the amplitude of oscillations,
and the dispersion relation. Thus 
the generalized susceptibility $\chi _{\alpha \beta }^{xx}$,
which is the response
function, is proportional to the inverse dynamical matrix 
\begin{equation}
\chi _{\alpha \beta }^{xx}(n,n^{\prime },\omega )\sim D_{\alpha \beta
}^{xx}(n,n^{\prime },\omega )^{-1}.
\end{equation}

We discuss here the cases
when the 
momentum transfer $\mathbf{Q}$ (parallel to the $x$ -axis)
lies close to 
the Bragg peaks of the composite crystal $ {\bf Q} = {\bf
Q}_{a, b}^{m, n} + {\bf q}$ with
\begin{eqnarray} 
\label{cur} Q_{a, b}^{m, n} = \frac{2\pi }{a}m+\frac{2\pi }{b}n 
\, , 
\end{eqnarray} 
where $m, n $ are integer numbers. Expression (\ref{cur}) implies that each
observed Bragg peak can be classified as follows: main reflections of the $a$ subsystem ($m\neq 0$,
$n=0$); main reflections of the $b$ subsystem ($m=0$, $n\neq 0$); common reflections to $a$ and $b$
subsystems ($m=0$, $n=0$); pure satellite reflections ($m\neq 0$, $n \neq 0$). With regard to the general
program we do not have the ambition to find the dynamical structure factor in the whole region of
parameters. In fact such a general formulation would not be tractable and in 
any case it would have
little meaning in view of the qualitative nature of our model. For this reason
and
only for illustration purposes
we consider a few particular cases.

Consider first the 
forward scattering when the momentum
transfer $\mathbf{Q}=\mathbf{q}$ is small
(acoustic case). This regime corresponds 
to Brillouin (light and X-ray) scattering. On the basis of (\ref{SQ}) the structure factor 
$S(Q,\omega )$ is reduced to (aiming at a qualitative description of the dynamics 
we drop the amplitude factor $(k^{\prime }/\pi k)\exp [-2W_{{\bf q}}]\left (1-\exp (\hbar \omega /T)\right
)^{-1}$, where $W_{\bf q}$ is the Debye - Waller factor):
\begin{equation}
S(q,\omega )\propto \sum _{\alpha \beta }(q^{2}/\omega )s_{\alpha }s_{\beta } Im D_{\alpha
\beta }^{xx}(q,\omega )^{-1}
\end{equation}
where $D_{\alpha \beta }^{xx}(q,\omega )$ is given by (\ref{matr}).

As mentioned earlier in order to pick 
out the effects of the dynamical dissipative coupling we neglect the
elastic coupling $\lambda ^{ab}$ and assume that both channels $a$ and $b$ 
are equally-participating in the scattering process ( $s_{a}=s_{b}=s$).
Then $S(q,\omega )$ can be written in a simple form: 
\begin{gather}
S(q,\omega )\propto \label{ac} \\
\frac{s^{2}(\gamma _{a}+\gamma _{b})(\rho _{a}-\rho
_{b})^{2}q^{2}[\omega ^{2}-c_{-}^{2}q^{2}]^{2}}{(\omega
^{2}-c_{a}^{2}q^{2})^{2}(\omega ^{2}-c_{b}^{2}q^{2})^{2}+(\gamma _{a}+\gamma
_{b})\omega ^{2}(\omega ^{2}-c_{+}^{2}q^{2})^{2}}\,,  \notag
\end{gather}
with:
\begin{equation}
c_{\pm }^{2}=\frac{(\rho _{a}c_{a}^{2}\pm \rho _{b}c_{b}^{2})}{\rho _{a}\pm
\rho _{b}}.
\end{equation}

The structure factor
in the so-called pseudo acoustic regime   
(when $\mathbf{Q}$ is close to a
nonzero Bragg peak) is different from the acoustic structure factor (\ref{ac}). 
It is worth noting also that this feature is different from the case of 
phonons in conventional
crystals. 

Consider e.g. the pseudo acoustic region close to the Bragg peak $Q_{1a}=\pi
/a$ (Fig. 1).  Because of the large energy separation, the modes $a$ and $b$
are decoupled and the structure
factor for mode $a$ is written as: 
\begin{equation}
S_a(\pi /a+q,\omega )\propto (Q_{1a}/\omega )s_{a}^{2}D_{a a}^{xx}(\pi /a+q,\omega )^{-1}.
\end{equation}
The matrix element $D_{a a}^{xx}(\pi /a+q,\omega )^{-1}$
coincides with $D_{a a }^{xx}(q,\omega )^{-1}$ (it corresponds
to the umklapp coupling processes).  On one side
the $"a"$ phonon with $Q=\pi /a+q$ is
decoupled from the $"b"$ phonon with $Q=\pi /a+q$ because of the
large energy separation. At the same time the $"a"$ phonon with $Q=\pi /a+q$ is
equivalent to the $"a"$ phonon with $q$, that in turn, is strongly coupled with 
the $"b"$ phonon with $Q=q$.

A straightforward calculation gives 
\begin{gather}
S_a(\frac{\pi }{a}+q,\omega )\propto \\
\frac{Q_{1a}^{2}(\gamma _{a}+\gamma
_{b})\rho _{a}^{2}(\omega ^{2}-c_{a}^{2}q^{2})^{2}}{(\omega
^{2}-c_{a}^{2}q^{2})^{2}(\omega ^{2}-c_{b}^{2}q^{2})^{2}+(\gamma
_{a}+\gamma _{b})\omega ^{2}(\omega ^{2}-c_{+}^{2}q^{2})^{2}\,}.  
\notag
\end{gather}
The results of the calculations of $S(\mathbf{q},\omega )$ for different
values of $\omega $ scanning the structure factor at a fixed value of $q$ 
are shown in Fig. 2.
\begin{figure}[tbp]
\centerline{\includegraphics[width=8.5cm]{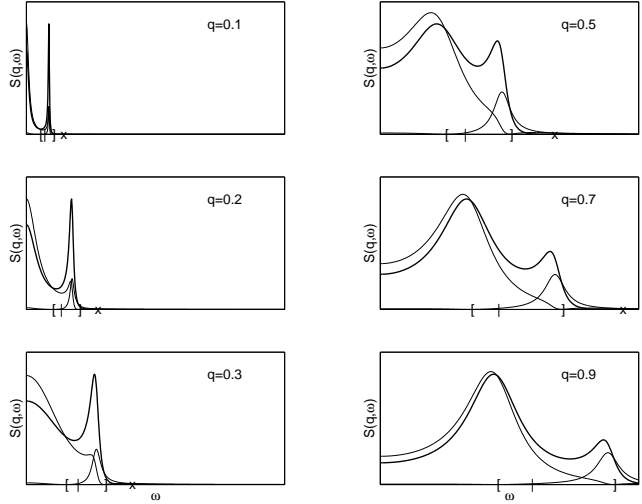}}
\caption{The structure factor of a uniaxial composite system. Bold lines correspond
to $S(q,\protect\omega )$, normal lines correspond to $S(Q_{a}^{m}+q,\protect\omega )$ and
$S(Q_{b}^{m}+q,\protect\omega )$. Ticks: ''['' and '']''
correspond to $c_{a}q$ and $c_{b}q$; ticks ''$|$'' and ''x'' correspond 
to $c_{+}q$ and $c_{-}q$. Parameters: $\protect\rho _{a}=1,$ $\protect\rho 
_{b}=2,$ $c_{a}=1$, $c_{b}=2$, $\protect{\gamma _{a} + \gamma _b} =0.45$.}
\label{S(qw)}
\end{figure}
The picture presented above leads
to the following qualitative predictions:
\begin{itemize}

\item

The maximum in $S({\bf q} , \omega )$ due to the sublattice
phonon with the lower frequency disappears at a certain wavevector. 

\item 

Both $S(q,\omega )$ and $S(n\pi /a+q , \omega )$, $S(m\pi /b +
q , \omega )$ have asymmetric profiles with ''antiresonance'' points where
they vanish (as in the case of the Fano resonance, when 
the interference of a discrete state with a continuum gives rise to
characteristically asymmetric peaks in excitation spectra); 

\item
Although $S(n\pi /a + q , \omega )$, $S(m\pi /b + q , \omega )$
correspond to ''decoupled'' phonons, they show traces of mixing between
the two phonon branches because of the umklapp process;

\item

At low $q$ we always have a double  peak structure with a central
peak and a sound-like side peak.

\end{itemize}

\section{V. Conclusion}

Let us sum up the results of our paper. We propose a simple phenomenological
model describing a composite crystal, constructed from two sets of periodic
inter-penetrating chains having a common direction (either commensurate or
incommensurate). 
We intentionally restrict ourselves to the simplest case taking into account 
mainly the dissipative coupling of the two subsystems, since theoretical
modelling of composite crystals is often hampered by lack of 
detailed knowledge concerning the parameters entering more complete
and more sophisticated models than the present one.

As incommensurate composites do not possess translational symmetry,
one cannot define a Brillouin zone. In zero order approximation if both sublattices are
independent, there are two phonon modes (one for each sublattice)
but only one 
sound velocity (in the long-wavelength limit). Coupling between sublattices
leads (again in the hydrodynamic, i.e. $q \to 0$ limit) to one acoustic mode
for the composite, and one sliding mode, as found above. In this limit the acoustic
mode is associated with a global translation of the composite, while the sliding
mode describes the relative ''dressed'' translation of the two sublattices.

We find the eigenmodes of the system and 
calculate the dynamical structure factor $S({\bf Q}, \omega )$.
We show that at high
frequencies there are two longitudinal sound modes with wave vector independent
attenuation, while in the low frequency region there is one propagating
sound mode and two over-damped phase modes.
The structure factor has a specific  profile resembling the Fano resonance.
We anticipate that these effects will be observable, and that understanding
the physical mechanism will be essential
to predict the behaviour of the composite materials.
 
However a number of remarks related to our results are in order. 
First, the separation of high and low frequency regions depends
on the natural scale $q_c \propto \gamma  /c$ (see Eqs. (\ref{x} - \ref{x3})), and the large
$q$ or $\omega $ limit corresponds to the uncoupled case, when both eigenvectors are
those of acoustic modes propagating more or less independently, while in the opposite
limit the chains are tightly coupled.
Of course in real materials the critical wave vector
$q_c$ could be so large that the coupled limit is adequate over the whole
accessible range of wave vectors.

Second, in fact we
considered a composite chain system embedded into two dimensional space.
Thus one should distinguish between modes propagating along the $x$ and $y$
directions. For the modes propagating along $x$ we found in the low $q$
limit two diffusive and one pair of degenerate sound like modes. 
In addition there are transverse modes which are: two time-conjugated 
optical ones (i.e. with a gap) and two time-conjugated acoustic modes (sound
like). For the propagation along $y$ the situation is inverse: transverse
modes give two diffusive modes and two time-conjugated sound like modes, while
longitudinal modes are two time-conjugated optical modes and two 
time-conjugated acoustic modes.

Note also that it is to be expected that the forces coupling the two chains are weak 
for relative displacements along the chain direction but of normal
strength for displacements perpendicular to the chains. It implies that even the
smallest wave vectors parallel to the $x$ - axis probed e.g. by neutrons
(see below our short survey of experimental data) may fall within the large
$q$ uncoupled regime, whereas all wave vectors in the orthogonal
directions are in the normal coupled regime.

Third, we assumed that atomic motions are not related to charge fluctuations
(therefore we neglected Coulomb forces). As a result of Coulomb interactions
additional $q=0$ gaps (plasmon gaps) will open, and in addition long-range
Coulomb interactions will lead to a renormalization of the short-range
elastic constants. The detailed discussion of all these problems is beyond
the scope of our paper. 

It is worth to compare 
the predictions of the present 
model, with results of \cite{FR82,FR83} where the authors also
stressed the relevance of dynamical couplings. Qualitatively
(especially in the low frequency region) our model leads to the same three
longitudinal modes as predicted in \cite{FR82,FR83}. 
However we presented a much simpler derivation which offers a deeper insight 
into the physical structure of the problem.
Besides we should stress
that we determined as well the dispersion laws for 
the transverse modes and
for the high-frequency region we found an unusual $q$-independent contribution
to the sound modes attenuation (due to our dynamical coupling) superimposed 
on the conventional ($\propto q^2$) line-width. Note also that
for our model there are no universal inequalities between sound 
mode velocities $c_a , c_b , c_s$ as reported in \cite{FR82,FR83}.

In spite of the current research activity in the field of composite crystals, 
there are still few experimental results to which model predictions can be compared. 
Furthermore, the generic aspects of composite crystals which this and other models aim 
to describe, are often masked in practice by other effects, specific to each particular 
compound under study. One of the first composite systems studied with respect to dynamics, 
is the $Hg_{3-\delta }AsF_6$ ''mercury-chain'' compound, which turned out to be a rather complex 
and unique system with two orthogonal sets of $Hg$ - chains, showing 
1d liquid-like behaviour at high temperature ($T>T_c = 120 K$) \cite{HA79}. 
Below $T_c$, the long-range phase-correlation between parallel $Hg$ - chains is established 
through indirect interchain interactions, mediated by the orthogonal set of $Hg$ - chains 
(which order simultaneously). Hence the compound is not a good 1d - composite model system.

The alkane-urea compounds \cite{exp2} appear a priori as better model systems 
because of their strongly uniaxial hexagonal structure. However, the structural transition 
at $T_c \simeq 130 K$, marking the onset of ferroelastic domains, 
precludes a detailed inelastic neutron or Brillouin study 
at low temperatures, where mode-broadening due to molecular disorder and 
anharmonic effects could be suppressed. 
The report by Schmicker et al., 
\cite{SV95} of Brillouin scattering evidence for a propagating 
phase mode, has clearly been refuted by subsequent measurements by Ollivier et al., \cite{OE98}. 
Instead, these latter authors report observing essentially normal acoustic dispersions 
and a quasielastic line polarized in the chain direction, 
which they ascribe to an overdamped, pinned phase mode. In 
the limit when the mode damping coefficient $f $ exceeds the pinning 
frequency $\omega _g$ ($\omega _g
\ll f $), the phase-mode response would become quasielastic with a 
width $\gamma _{Q.E.} = \omega _g^2/f $. 
It is indeed quite plausible that, in the $q$-range covered 
by light scattering, neither $\omega _g$ nor $f $ would show much 
$q$-dependence, in which case one
would not 
expect much $q$-dependence for the quasielastic linewidth $\gamma _{Q.E}$, 
in agreement with the results
in ref. \cite{OE98}. 
It conforms as well with our theoretical predictions presented in Section III.
The fact that $\gamma _{Q.E}$ 
is also found to be $T$-independent between 
$130 K$ and $300 K$, is much more puzzling since, over such a wide $T$-range, 
one would expect $\omega _g$ to soften and $\Gamma $ to increase with $T$, 
leading to a significantly narrower quasielastic line at room-$T$, 
in clear contradiction with the measurements. Hence, the assignment of the 
observed quasielastic line as a collective excitation must be taken as tentative only. 
On the other hand inelastic neutron scattering data on the n=19 alkane-urea compound 
\cite{thessis}, 
show only one longitudinal acoustic (LA) mode in the chain direction 
(instead of two as would be expected in the high-q limit), with an 
anomalous q-independent linewidth. The non-observation of the collective 
modes associated with the alkane sublattice can be ascribed to the strong orientational 
and translational disorder of the alkane molecules \cite{GS90}, \cite{FU94} above 
$T_c$.

Another interesting system is the aperiodic layered crystal $Bi_2Sr_2CaCu_2O_{8+\delta }$, 
which was recently suggested to belong to the class of composite 
(rather than modulated) incommensurate systems \cite{EB00}, \cite{EB01}. 
The mode structure observed by Etrillard et al. \cite{EB01} is consistent with 
the presence of two independent $LA$ branches along the incommensurate direction, 
as expected in the $q$-range of the neutron experiment. 
The relative strength of the two branches is however not understood, nor is the 
absence of a similar sublattice decoupling for the $TA$ branches polarised in the incommensurate
direction.

It is obvious from the above remarks that the experimental situation still remains very open. 
To a large extent, this is due to the difficulty in identifying suitable model 
systems and in growing single-crystal specimens of appropriate size 
and quality for light and neutron scattering experiments. We hope that this work stimulates 
further efforts in this field. 

\acknowledgements 
One of the authors (E.K) is indebted to 
RFFR and INTAS Grants for parial support. 

\section{Appendix}

The equilibrium positions of the atoms in the incommensurate direction
($x$) could be written in a more abridged form than (\ref{ckl1}), namely
$$ 
x_{n}^{a}(t) = na +u_x^a + 
g_1(na + u_x^a - u_x^b) \, ;
$$
$$
x_{m}^{b}(t) = mb +u_x^b + g_2(mb + u_x^b - u_x^a) \, ,
$$
with $g_1(x + b) \equiv g_1(x)$ and $g_2(x+a) \equiv g_2(x)$.
Both functions $g_1$ and $g_2$ can be found from a static microscopic model.

The energy of interaction within the chains is written as usually, the dynamical coupling in the
dissipation function can be written as: 
\[ R=\frac{1}{2}\Gamma
\sum_{mn}(\stackrel{.}{u}_{n}^{a}-\stackrel{.}{u} _{m}^{b})^{2}f(ma-nb) 
\, ,
\] 
where $f(x)$ can be
modelled as: 
\[ 
f(x)= \exp \left [- \frac{x^2}{\lambda ^2} \right ] \, . 
\] 
Suppose we have a relatively long-range coupling
$\lambda \gg a, b $. In the case similar to that considered in the main text of the paper 
the equations of
motion (after Fourier transformation over $t$) read 
$$ 
-\omega ^{2}\rho _{a}u_{n}^{a}
=\lambda _{a}(u_{n+1}^{a}+u_{n-1}^{a}-2u_{n}^{a})+
$$
$$
i\Gamma \sum_{m}f(ma-nb)u_{m}^{b}\, ;
$$
$$ 
-\omega ^{2}\rho
_{b}u_{n}^{b} =\lambda _{b}(u_{n+1}^{b}+u_{n-1}^{b}-2u_{n}^{b})
$$
$$
+i\Gamma \sum_{m}f(ma-nb)u_{m}^{a} \, ,
$$
or, after space Fourier transformation in $ma$, $nb$: $u_{n}^{a}=u_{n0}^{a}\exp {i(\omega
t-qna)}$, $u_{n}^{a}=u_{n0}^{b}\exp {i(\omega t-qnb)}$ 
$$
-\omega ^{2}\rho _{a}u_{n0}^{a}
=-\lambda _{a}\sin ^{2}\frac{qa}{2} u_{n0}^{a}+
$$
$$
i\Gamma \sum_{m}e^{-(ma-nb)^{2}/\lambda
^{2}}e^{iq(ma-nb)}u_{m0}^{b} \,;
$$
$$ 
-\omega ^{2}\rho _{b}u_{n0}^{b} =-\lambda _{a}\sin ^{2}\frac{qb}{2}
u_{n0}^{b}
$$
$$
+i\Gamma \sum_{m}e^{-(ma-nb)^{2}/\lambda ^{2}}e^{iq(ma-nb)}u_{m0}^{a} 
$$ 
These equations give $\chi _{\alpha \beta }^{xx}(n,m,\omega )$ 
which we need for the analysis of
the neutron scattering.

\end{document}